\documentclass[prb,twocolumn]{revtex4}
\usepackage{russian}
\begin{document}
\title{SUPERCONDUCTING STATE OF EXCITONIC INSULATOR}
\author{E. G. Batyev} \address{Institute of Semiconductor Physics,
Siberian Branch of Russian Academy of Sciences, Novosibirsk,
630090, Russia}\thanks{e-mail:batyev@isp.nsc.ru}

\begin{abstract} A state of an excitonic insulator with the electric
current is studied. Initially, in the metallic phase, the
electrons and holes are assumed to be moving in the opposite
directions, so as the electric current exists. This state is
supported by an external condition (the specimen is in an electric
circuit with the current). When the temperature decreases, the
transition to the ordering state due to formation of the
electron--hole pairs is possible (similar to the ordinary state of
the excitonic insulator). The properties of the state at zero
temperature are investigated. The spectrum of elementary
excitations has a gap, and so the conclusion can be made that
obtained state is superconducting one. Thus, depending on the
external conditions, excitonic insulator behaves itself like the
insulator or superconductor. That is correct in the limit of
strong overlapping of the electron--hole pairs.\end{abstract}
\maketitle PACS: 74.10.+v; 71.35.-y; 71.30.+h

It was shown in the work \cite{1} that, in the system of electrons
and holes, the Cooper effect \cite{2} is possible due to Coulomb
attraction of electrons and holes, and the formation of
Bose--condensate of the electron--hole pairs takes place (as in
the case of Cooper pairs for superconductors \cite{3}). For the
case without the so--called phase fixation, the superfluid motion
is possible but, for the system with equal amounts of electrons
and holes, there is no electric current for this motion. Therefore
it is said on the excitonic insulator rather than superconductor.
Nevertheless, it is possible to utilize the superfluidity for the
system of spatially separated electrons and holes. This idea was
put forward in the work \cite{4}.

Interest to such a system is mainly supported because the
temperature interval of the ordering state may be greater than for
superconductors. It is especially important for the bilayer
quantum well systems \cite{4}.

Interaction of the electric charges in excitonic insulator is
described with the help of dielectric constant (which is large
\cite{5}) just as in ordinary insulators, and the same is for a
response to alternating electric field \cite{6}. The impression
arises that the excitonic insulator behaves entirely as the
ordinary dielectric (insulator). However, some difference is
possible. In order to make sure in that, one can bring the
following argument.

At first, we use a simplified approach and imply that the states
of noninteracting electrons and holes (close to extrema of energy
zones) may be described by the plane waves with the corresponding
effective masses as usual. That is to say, the Hamiltonian $H$ of
the system is \begin{eqnarray} \nonumber H = \int d{\bf r}
\Psi_e^+({\bf r})\biggl\{\frac{-\triangle}{2m_e}
-\mu_e\biggr\}\Psi_e({\bf r}) +\\ \label{B} +\int d{\bf r}
\Psi_h^+({\bf r})\biggl\{\frac{-\triangle}{2m_h}-\mu_h\biggr\}\Psi_h({\bf r})+ \\
\nonumber +H_{e-h}+H_{e-e}+H_{h-h}\ .\end{eqnarray} Here $\hbar
=1$, $\Psi_e({\bf r})$ is the operator of the electron field,
$m_e$ is the effective mass of the electron, $\mu_e$ is the
chemical potential of electrons which is counted from the
appropriate energy minimum (the index $h$ corresponds to the
holes). The other terms mark the interactions, for example,
$H_{e-h}$ is the interaction of the electrons with the holes.

Let the ground state wave function of an excitonic insulator is
$\Phi_0({\bf R}_e,{\bf R}_h)$ where the ${\bf R}_e$ and ${\bf
R}_h$ are the sets of the electron and hole coordinates. Let us
consider the trial wave function $\Phi$ which differs from
$\Phi_0$ only by the phase factors, namely
\begin{eqnarray} \nonumber \Phi({\bf R}_e,{\bf R}_h) = \Phi_0({\bf R}_e,{\bf R}_h)\times \\
\label{1} \times \exp\biggl\{ i{\bf p}_e\sum_n {\bf r}_n +i{\bf
p}_h\sum_{n'} {\bf r}_{n'} \biggr\}\end{eqnarray} (the sums are
over the electron and hole coordinates). The phase factors
correspond to the independent motions of the electron and hole
subsystems.

In order to find the corresponding mean energy $E$ of the system
for the state (\ref{1}), it is enough to determine the one-- and
two--particle density matrices. For example, the one--particle
density matrix for the electrons is
$$\rho({\bf r',r}) = <\Psi^+_e({\bf r'})\Psi_e({\bf r})>\ ,$$
$$\Psi_e({\bf r}) = \frac{1}{\sqrt{V}}\sum_{\bf p}a_{\bf
p}\exp(i{\bf pr})\ .$$ Here $V$ is the volume of the system (the
area in two--dimensional case), and the symbol $<...>$ denotes the
averaging over the state of the system. One can see that
\begin{equation} \label{C} \rho({\bf r',r}) = \rho_e({\bf
r',r})\exp\Bigl[i{\bf p}_e({\bf r-r'})\Bigr]\end{equation}
($\rho_e$ corresponds to the ground state of motionless
subsystems). The similar phase factors appear in the two--particle
density matrix. Therefore, the mean value of potential
interactions equals to the value for the motionless subsystems
(including the ordering energy). But the kinetic energy changes,
and we have $$\frac{E-E_0}{V} =\frac{p_e^2 }{2m_e}\ n_e
+\frac{p_h^2}{2m_h}\ n_h \ ,$$ where $n_{e,h}$ are the
concentrations of electrons and holes, $E_0$ is the ground--state
energy of motionless subsystems. If the electric charges of
electrons and holes are designated by $\pm e$ then the density of
electric current can be written as $${\bf j} = \frac{e{\bf
p}_e}{m_e}\ n_e-\frac{e{\bf p}_h}{m_h}\ n_h\ .$$ For a given value
of the current, the total energy is minimal in the case:
\begin{equation} \label{A}{\bf p}_e =-{\bf p}_h \equiv{\bf p}_0\
.\end{equation}

\textbf{The model and results.} The arguments cited above show the
direction of search only. It is necessary to consider the problem
in detail. Let in the metallic phase, when the temperature is
higher than the point of the transition, there is a current with
electrons and holes moving in opposite directions. Under lowering
of temperature, because of Coulomb attraction between electrons
and holes, the formation of electron--hole pairs (of Cooper type)
is possible just as for motionless subsystems. As a result, the
state with a gap in quasiparticle spectrum arise likewise for the
ordinary excitonic insulator but with electric current.
Apparently, this current is superconducting.

We shall consider the problem by the simplest way. Firstly, we use
a model with one type of electrons (corresponding operators of
creation and annihilation are $a^+_{\bf p}$ and $a_{\bf p}$) and
with one type of holes (corresponding operators of creation and
annihilation are $b^+_{\bf p}$ and $b_{\bf p}$). The spectra and
concentrations of electrons and holes are assumed to be the same
($m_e=m_h\equiv m,\ n_e=n_h\equiv n$). Secondly, we introduce an
additional condition which takes into account the motions of the
electron and hole subsystems. And thirdly, we simplify the model
exactly by the same way as in the theory of
Bardeen--Cooper--Schrieffer (BCS) \cite{3}.

The Hamiltonian of the system is
\begin{eqnarray}\nonumber H =\sum_{\bf p} \xi({\bf
p})\biggl\{a^+_{\bf p}a_{\bf p} + b^+_{\bf p}b_{\bf p}\biggr\}+\\
\label{2} +\ \frac{1}{V}\sum W({\bf p}_1-{\bf p}_4)
a^+_1b^+_2b_3a_4 +\end{eqnarray} $$+H_{e-e}+H_{h-h}\ .$$ Here
$\xi({\bf p})$ is the energy of particle counted from the Fermi
energy, and the interaction operator $H_{e-h}$ is written down
explicitly (see (\ref{B})). It is necessary to emphasize that, in
two--dimensional case, when the electrons and holes are separated
by a barrier, its zones may be displaced in energies (this
difference may be supported by an external potential between the
electron--hole layers).

For the moving subsystems, we add an appropriate condition so as,
instead of Hamiltonian (\ref{2}), we have:
\begin{eqnarray}\nonumber H({\bf v}_{e,h}) =\sum_{\bf p}\biggl\{
\xi_e({\bf p})a^+_{\bf p}a_{\bf p} + \xi_h({\bf p})b^+_{\bf
p}b_{\bf p}\biggr\}+\\ \label{3} +\ \frac{1}{V}\sum W({\bf
p}_1-{\bf p}_4) a^+_1b^+_2b_3a_4+H_{e-e}+H_{h-h}\ ,\end{eqnarray}
$$\xi_{e,h}({\bf p}) = \xi({\bf p})-{\bf pv}_{e,h} +
\frac{m\textrm{v}_{e,h}^2}{2} =\xi({\bf p}-m{\bf v}_{e,h}) \ .$$
Here $\bf p$ is the momentum of particle in the static frame of
reference, ${\bf v}_{e,h}$ are the velocities of the corresponding
subsystems. For zero temperature, the one-particle states with
negative energies are filled only.

For arbitrary velocities, the Bose--condensate of electron--hole
pairs with nonzero momentum arises. When  $m{\bf v}_e=-m{\bf v}_h
\equiv {\bf p}_0$ (see (\ref{A})), then the pair momentum equals
to zero by symmetry. Further we consider just the case, and
therefore one can write: \begin{equation}\label{4} \xi_e({\bf p})
= \xi_h({\bf -p}) =\xi({\bf p} - {\bf p}_0)\equiv\tilde{\xi}({\bf
p})\ .\end{equation}

As a result, we have exactly the same problem as for a
superconductor. Therefore one may use the model of BCS type
\cite{3}, i. e. by taking into account only the part of the
interaction which is responsible for the formation of the
electron--hole pairs with zero momentum. After that, instead of
(\ref{3}) and bearing in mind of (\ref{4}), we have:
\begin{eqnarray}\nonumber H({\bf v}_{e,h}) \rightarrow H({\bf
p}_0) = \sum_{\bf p}\tilde{\xi}({\bf
p})\Bigl\{a^+_{\bf p}a_{\bf p} + b^+_{\bf -p}b_{\bf -p}\Bigr\}+\\
\label{5} +\ \frac{1}{V}\sum_{\bf p,p'} W({\bf p}-{\bf p}')
a^+_{\bf p}b^+_{\bf -p}b_{\bf -p'}a_{\bf p'}\ .\end{eqnarray} That
is our model Hamiltonian.

The well--known self-consistent (mean--field) approximation fits
ideally for analysis of the model Hamiltonian (\ref{5}). In this
approximation, the interaction operator $H_i$ (the last term in
(\ref{5})) is written as \begin{eqnarray} \nonumber H_i
\rightarrow \sum_{\bf p}\Bigl\{
a^+_{\bf p}b^+_{\bf -p}\Delta({\bf p}) + H.c. \Bigr\}- \\
\label{6}-\frac{1}{V}\sum_{\bf p,p'} W({\bf p}-{\bf
p}') <a^+_{\bf p}b^+_{\bf -p}><b_{\bf -p'}a_{\bf p'}>\ ,\\
\nonumber \Delta({\bf p}) \equiv\frac{1}{V} \sum_{\bf p'} W({\bf
p-p'}) <b_{\bf -p'}a_{\bf p'}>\ .\end{eqnarray} Here the symbol
$<...>$ corresponds to averaging over the state of the system.

It is necessary to emphasize that, in this case, the
self-consistent approximation gives asymptotically (i. e. in the
limit $V\rightarrow\infty$) correct result. It is clear from the
pseudo--spin approach of Anderson \cite{7} since, in this
approach, an every pseudo--spin interacts with all others while
its number is macroscopically large so that the fluctuations are
not essential. (We remind that the pseudo--spin $1/2$ is
introduced for the every point of the momentum space and the up
and down pseudo--spin projections correspond to the filled and
empty states of the electron--hole pair in the point.)

Let us come back to the Hamiltonian (\ref{6}). The part $h$ of
this operator corresponding to the pair of particles with momenta
$\pm{\bf p}$ has the form: \begin{equation} \label{7} h =
\tilde{\xi}(a^+a + b^+b) +\Delta (a^+b^+ + ba)\ .\end{equation}
That is for the real order parameter $\Delta$ (the indices are
omitted). This operator can be reduced to diagonal form by
Bogolyubov transformations, namely:
\begin{eqnarray}\nonumber a=u\alpha +v\beta^+\ ,\ \ \ \ b= u\beta -v\alpha^+\
;\\ (u^2,\ v^2)= \frac{1}{2}\biggl\{1\pm\frac{\tilde{\xi}}
{\epsilon} \biggr\}\ ,\ \ \ \ \ uv=-\frac{\Delta}{2\epsilon}\ ;
\label{8}\\ \nonumber \epsilon\rightarrow\epsilon({\bf p})
=\sqrt{\tilde{\xi}^2({\bf p}) +\Delta^2({\bf p})}\ .
\end{eqnarray} Here $\alpha$ and $\beta$ are quasi--particle operators of
Fermi type and $\epsilon({\bf p})$ is the quasi--particle energy.
As a result, the operator $h$ is rewritten in the form:
\begin{equation}\label{9} h \rightarrow \epsilon({\bf
p})(\alpha^+_{\bf p}\alpha_{\bf p} + \beta^+_{-\bf p}\beta_{-\bf
p}) + \Bigl[\tilde{\xi}({\bf p})-\epsilon({\bf p})\Bigr]\
.\end{equation}

The equation for the order parameter (\ref{6}) at zero temperature
has the form: \begin{equation}\label{10} \Delta({\bf p}) = -\
\frac{1}{V}\sum_{\bf p'} W({\bf p-p'})\ \frac{ \Delta({\bf
p}')}{2\ \epsilon({\bf p'})}\ .\end{equation} When one recall the
definition of $\tilde{\xi}$ (\ref{4}) then one can see that our
order parameter is expressed via the order parameter for
motionless subsystems $\Delta_0({\bf p})$ (under the condition
${\bf p}_0=0$) by the following way: \begin{equation}\label{11}
\Delta({\bf p}+{\bf p}_0) = \Delta_0({\bf p})\ .\end{equation} The
same holds for the spectrum of quasi--particles.

The obtained state entirely corresponds to the trial function
(\ref{1}). In order to make sure in that, it is enough to
determine, for example, the one--particle density matrix
$\rho({\bf r',r})$ of electrons: $$\rho({\bf r',r}) =
<\Psi^+_e({\bf r'})\Psi_e({\bf r})>\ .$$ By passing to the
quasi--particle operators (\ref{8}) and doing the corresponding
calculations, we receive the previous result (\ref{C}) exactly.
The same result can be obtained by using the function of BCS type
for our case: \begin{equation}\label{12}\Phi = \prod_{\bf
p}\biggl\{u({\bf p}) + v({\bf p})a^+_{\bf p}b^+_{\bf -p}
\biggr\}|0> \ .\end{equation} The other conclusions are the same
as for the trial function (\ref{1}).

It is necessary to note that the function (\ref{12}) gives the
minimal mean value of the operator (\ref{5}), i. e. it corresponds
to minimum of the energy of the system at the given motions of the
subsystems. When our specimen is inserted in an electric circuit
with a current then, strictly speaking, only the total current of
the system is determined by the external conditions but not the
current of every subsystem. However, under given total current,
the zero momentum of electron--hole pair corresponds to the
minimal energy of the system (see (\ref{A})) and  just the same
was supposed. As long as there is the gap in the excitation
spectrum, this current, apparently, flows without resistance.

\textbf{Discussion.} 1) We start with the trial function
(\ref{1}). This function gives a growth of the total energy
through the kinetic energies of the subsystems only, and the
interaction energy is the same as for the ground state (including
the energy of ordering). Formally, such a function can be written
for any system but there is no point in doing that in any case.
For our problem, it is reasonable only for the case of strong
overlapping of electron--hole pairs. In that case, we have the
correlated pairs in such a way that, for an electron with some
momentum, there is a hole with opposite momentum. That takes place
both for the motionless and for the moving subsystems. And so we
may use the same model (\ref{5}) for both cases. It is obvious
that the model (\ref{5}) and the corresponding conclusions are not
correct for sufficiently small size of the pair, i. e. if the pair
size is smaller than the average distance between the pairs.
Indeed, one can imagine a counter--flow of the electrons and holes
for continuous medium but can not do it for the individual
electron--hole pairs. Apparently, under lowering of
concentrations, there is a transition between two regimes, namely,
between high concentrations with possibility of nondissipative
current (our case) and low concentrations with the insulator
properties only.

2) Our state with nondissipative (superconducting) current is
supported by an external 'force' -- by a current in an electric
circuit in which our specimen is included. Therefore, we
considered the model with moving subsystems. But when there is no
external 'force' then our state become a time-dependent one. Let
us imagine that our state is created in an initial point of time
and then it is left by itself. We can take our function (\ref{12})
as initial function and can expand it over the proper states of
motionless subsystems. As a result, for the electron current ${\bf
j}_e(t)$, we find: \begin{equation} \label{13} {\bf j}_e(t) =
\frac{en{\bf p}_0}{m}\ \int_0^\infty\frac{\cos(\sqrt{x^2 +1}\
\tau)}{(x^2 +1)^{3/2}}\ dx \end{equation} $$(\tau = 2\Delta_0t)$$
(and the same result takes place for the holes). This is natural
because, in our system without impurities, only total momentum is
conserved but not the difference of the momenta of electrons and
holes. Apropos, for the state (\ref{12}), the fluctuations of the
current are relatively small ($\sim 1/\sqrt{V}$).

3) A restriction, that spectra of electrons and holes are the
same, is not essential. But it is necessary the concentrations of
electrons and holes to be sufficiently close. It is known from the
theory of superconductivity that, when the difference of
concentrations increases, then at first a periodic order state
arises and next the ordering disappears at all \cite{8}.

4) Another restriction can arise from collisions with impurities.
In general, the electrons and holes collide differently with
impurities. Therefore, the ordering can vanish for the
sufficiently large difference, initially a gapless regime arising
\cite{9}.

5) It is known that the phase fixation removes the superfluidity
in excitonic insulators \cite{10}. But in two--dimensional case,
when electrons and holes are separated by sufficiently large
barrier, it is insignificant \cite{4}. One can expect that, for
our state, the derived results are insensitive to cited effect (in
two--dimensional case).

In conclusion, there is good reason to believe that, for an
excitonic insulator, an electric current without dissipation is
possible. That is shown by using of a simple model. Probably, this
state can be observed experimentally in a quasi two--dimensional
case when electrons and holes are separated by a barrier (for the
bilayer quantum well systems \cite{4}, see also the paper
\cite{11}).

I thank A. V. Chaplik and M. V. Entin for discussion.

This work is supported by the Russian Foundation for Basic
Researches(Grant 06--02--16923) and by the Council for Supporting
of Leading Scientific Schools under President of the Russian
Federation (Grant NSh--4500.2006.2).


\begin{thebibliography}{99}
\bibitem{1} L. V. Keldysh and Yu. V. Kopaev, Fiz. Tverd. Tela \textbf{6}, 2791 (1964).
\bibitem{2} L. N. Cooper, Phys. Rev. \textbf{104}, 1189 (1956).
\bibitem{3} J. Bardeen, L. N. Cooper, and J. R. Schrieffer, Phys.
Rev. \textbf{108}, 1175 (1957).
\bibitem{4} Yu. E. Lozovik and V. I. Yudson, Zh. Eksp. Teor. Fiz. \textbf{71},738 (1976);
Sov. Phys. JETP \textbf{44}, 389 (1976).
\bibitem{5} E. V. Baklanov and A. V. Chaplik, Fiz. Tverd. Tela \textbf{7}, 2768 (1965).
\bibitem{6} E. G. Batyev and V. A. Borisyuk, Zh. Eksp. Teor. Fiz. \textbf{80}, 262 (1981).
\bibitem{7} P. Anderson, Phys. Rev. \textbf{112}, 1900 (1959).
\bibitem{8} A. I. Larkin and Yu. N. Ovchinnikov, Zh. Eksp. Teor. Fiz. \textbf{47}, 1137
(1964).
\bibitem{9} A. A. Abrikosov and L. P. Gor'kov, Zh. Eksp. Teor. Fiz. \textbf{39}, 1781
(1960).

\bibitem{10} R. R. Giseinov and L. V. Keldysh, Zh. Eksp. Teor. Fiz. \textbf{63}, 2255
(1972).
\bibitem{11} J. P. Eisenstein and A. H. MacDonald, Nature
\textbf{432}, 691 (2004).

\end{thebibliography}
\end{document}